\documentclass{article}
\usepackage{spconf,amsmath,graphicx}

\usepackage{cite}
\usepackage{caption}
\usepackage{subcaption}
\usepackage[]{algorithm}
\usepackage{algpseudocode}

\usepackage{amssymb}
\usepackage{verbatim}
\usepackage{url}
\name{Amir Adler$^{\star \dagger}$ \qquad Mati Wax$^{}$ \qquad Dimitrios Pantazis$^{\dagger}$}
  
  \address{$^{\star}$ Braude College of Engineering  \\
      $^{\dagger}$McGovern Institute for Brain Research, MIT}
\begin{document}

%
\title{Brain Source Localization by Alternating Projection}
\maketitle
\thispagestyle{empty}

\begin{abstract}
We present a novel solution to the problem of localizing magnetoencephalography (MEG) and electroencephalography (EEG) brain signals. The solution is sequential and iterative, and is based on minimizing the least-squares criterion by the Alternating Projection algorithm. Results from simulated and experimental MEG data from a human subject demonstrated robust performance, with consistently superior localization accuracy than scanning methods belonging to the beamformer and  multiple-signal classification (MUSIC) families. Importantly, the proposed solution is more robust to forward model errors resulting from head rotations and translations, with a significant advantage in highly correlated sources.
 
\end{abstract}


\section{Introduction}
Brain signals obtained by an array of magnetic or electric sensors using MEG or EEG offer the potential to investigate the complex spatiotemporal activity of the human brain. Localization of MEG and EEG sources has gained much interest in recent years since it can reveal the origins of neural signals during both normal and pathological brain function \cite{ilmoniemi2019brain}. 

Mathematically, the localization problem can be cast as  finding the location and moment  of the set of dipoles whose field best matches the MEG/EEG measurements in a least-squares (LS) sense \cite{mosher_multiple_1992}. In this paper we focus on \textit{dipole fitting} and \textit{scanning} methods, which solve for a small parsimonious set of dipoles and avoid the ill-posedness associated with imaging methods, such as minimum-norm \cite{hamalainen_magnetoencephalographytheory_1993}. 

The dipole fitting methods solve the optimization problem directly using techniques such as gradient descent, Nedler-Meade simplex algorithm, multistart, genetic algorithm, and simulated annealing \cite{uutela_global_1998,jiang_comparative_2003}. However, these techniques remain unpopular due to either  convergence to sub-optimal solutions or high computational complexity.

The scanning methods use a localizer function  and find  the dipole positions by  searching for the local maxima of the localizer function  throughout a discrete grid representing the source space \cite{darvas_mapping_2004}. The most common scanning methods are beamformers \cite{LCMV,VerbaRobinson} and MUSIC \cite{mosher_multiple_1992}, both widely used for bioelectromagnetic source localization.  Yet, when the correlation between the sources is significant, these method result in partial or complete cancellation of the correlated sources and to distorted  estimate of the time courses. 
The scanning methods are categorized as  \textit{non-recursive} or \textit{recursive}. The original Beamformer \cite{LCMV} and MUSIC \cite{mosher_multiple_1992}  are non-recursive,  and the localization  of multiple sources  require the identification of the largest local maxima in the localizer function. Some multi-source variants are also non-recursive (e.g. \cite{hui_identifying_2010, CorrBF, DCBF, EDCBF}), and  they either use brute-force optimization, assuming that the approximate locations of the  sources have been identified a priori, or still require the identification of the largest local maxima in the localizer function. To overcome these limitations,  recursive counterparts of the non-recursive methods which search for one source at a time, such as RAP-MUSIC \cite{Mosher1999}, Truncated RAP-MUSIC \cite{makela_truncated_2018}, and RAP Beamformer \cite{ilmoniemi2019brain}, have been developed. While recursive methods generally perform better than their non-recursive counterparts, they still suffer from several limitations, including limited performance, the need for high signal-to-noise ratio (SNR), or inaccurate estimation as correlation values increase.

 This paper presents a novel solution which overcomes these limitations.  Our starting point is the LS estimation criterion for the dipole fitting problem. This criterion yields a multi-source nonlinear and nonconvex minimization problem, making it very challenging to avoid being trapped in undesirable local minima \cite{SPMBrainMap}. To overcome this challenge, we propose the use of the Alternating Projection (AP) algorithm \cite{AP}. The AP algorithm transforms the multi-source problem to an iterative process involving only single-source maximization problems, which are computationally much simpler. Moreover, the algorithm has a very effective initialization scheme which is the key to its good  convergence. Here we demonstrate the robustness of the AP algorithm with MEG data, but the method is directly applicable to EEG.
\section{Problem Formulation}

In this section we briefly review the notations used to describe measurement data, forward matrix, and sources, and formulate the problem of estimating current dipoles. Consider an array of $M$ MEG or EEG sensors that measures data from a finite number $Q$ of equivalent current dipole (ECD) sources emitting signals $\{s_q(t)\}^{Q}_{q=1}$ at locations $\{\mathbf p_q\}^{Q}_{q=1}$. Under these assumptions, the $M\times 1$ vector of the received signals by the array is given by:
\vskip -10pt
\begin{equation}
\mathbf y(t) = \sum_{q=1}^{Q} \mathbf l(\mathbf p_q)  s_{q}(t) +\mathbf n(t),
\label{eq:snapshot1}
\end{equation}
where  $\mathbf l(\mathbf p_q)$ is the topography of the dipole at location $\mathbf p_q$  and $\mathbf n(t)$  is the additive noise. The topography $\mathbf l(\mathbf p_q)$, is given by: 
\vskip -5pt
\begin{equation}
\label{topography_model}
\mathbf l(\mathbf p_q) = \mathbf L(\mathbf p_q)  \mathbf{q},
\end{equation}
where $\mathbf L (\mathbf p_q)$ is the $M\times 3$ forward matrix at location $\mathbf p_q$ and $\mathbf q$ is the $3\times 1$ vector of the orientation of the ECD source. Depending on the problem, the orientation $\mathbf q$ may be  known, referred to as \textit{fixed-oriented} dipole, or it may be  unknown, referred to as \textit{ freely-oriented} dipole. Assuming that the array is sampled  $N$ times at $t_1,...,t_N$, the matrix $\mathbf Y$ of the sampled signals can be expressed as:
\vskip -10pt
\begin{equation}
\label{basic_equation}
\mathbf Y =\mathbf A(\mathbf P) \mathbf S +\mathbf N, 
\end{equation} 
where  $\mathbf Y$ is the $M\times N$ matrix of the received signals:
\vskip -10pt
\begin{equation}
\mathbf Y= [\mathbf y(t_1), ..., \mathbf y (t_N)],
\label{eq:snapshot4}
\end{equation}
$\mathbf A(\mathbf P)$ is the $M\times Q$ mixing matrix of the topography vectors at the $Q$ locations $\mathbf P=[\mathbf p_1,...,\mathbf p_Q]$:
\begin{equation}
\mathbf A(\mathbf P)=[\mathbf l(\mathbf p_1), ..., \mathbf l(\mathbf p_Q)],
\label{eq:snapshot5}
\end{equation}
$\mathbf S$ is the $Q\times N$ matrix of the sources:
\begin{equation}
\mathbf S= [\mathbf s(t_1), ..., \mathbf s (t_N)],
\label{eq:snapshot6}
\end{equation}
with $\mathbf s(t)=[s_1(t),..., s_Q(t)]^{T}$, and $\mathbf N$ is the $M\times N$ matrix of noise. We further make the following assumptions regarding  the emitted signals and the propagation model:

A1:  The number of sources $Q$ is known and $Q<M$.

A2:  The emitted signals are  unknown and  arbitrarily correlated, including  the case that a subset of the sources or all of them are synchronous. 

A3: The forward matrix $\mathbf L(\mathbf p)$ is known for every location $\mathbf p$ (computed by the forward model). 

A4: Every  $Q$ topography vectors $\{\mathbf l(\mathbf p_q)\}_{q=1}^Q$ are linearly independent, i.e., $\mathrm{rank}\mathbf A(\mathbf P)=Q$.

\hfill
\hfill

We can now state the problem of localization of brain signals as follows: \textit{Given the received data $\mathbf Y$, estimate the $Q$ locations of the sources $\{\mathbf p_q\}_{q=1}^{Q}$}.

\section{The Alternating Projection solution } 
We next present the solution for  fixed-oriented dipoles. The least-squares  criterion for this problem, from \eqref{basic_equation},    is  given by 

\begin{equation}
\label{LS_criterion}
\{\hat{\mathbf P},\hat{\mathbf S}\}_\text{ls} = \arg\min_{\mathbf P, \mathbf S}    \|\ \mathbf Y-\mathbf A(\mathbf P)  \mathbf S \|^{2}_{F}.
\end{equation}
where subscript $F$ denotes the Frobenius norm. To solve this minimization problem, we first eliminate the unknown signal matrix $\mathbf S$ by expressing it in terms of  $\mathbf P$. To this end, we equate   the derivative of \eqref{LS_criterion} to zero with respect to $\mathbf S$, and solve for $\mathbf S$:
\begin{equation}
\label{s_hat}
\mathbf{\hat  S} =  \mathbf (\mathbf A(\mathbf P)^{T}\mathbf A(\mathbf P) )^{-1}\mathbf A(\mathbf P)^{T}  \mathbf Y.
\end{equation}
Defining $\mathbf{\Pi_{\mathbf A(\mathbf P)}} $  the projection matrix on the column span of $\mathbf A( \mathbf P)$:
\vskip -15pt
\begin{equation}
\label{Projection_A}
\mathbf \Pi_{\mathbf A(\mathbf P)}=\mathbf A(\mathbf P)(\mathbf A(\mathbf P)^T\mathbf A(\mathbf P))^{-1}\mathbf A(\mathbf P)^T,
\end{equation}
equations \eqref{LS_criterion}, \eqref{s_hat}, and \eqref{Projection_A}, using  the rotation property of the trace operator and the idempotence property of the projection operator, yield:
\begin{eqnarray}
{\mathbf{\hat  P}} & = & \arg\min_{\mathbf P}  \|\ (\mathbf I -  \mathbf{\Pi}_{\mathbf A(\mathbf P)}) \mathbf Y \|^{2}_{F} \\
& = & \arg\max_{\mathbf P}  \|  \mathbf {\Pi}_{\mathbf A(\mathbf P)} \mathbf Y \|^{2}_{F} \\
& = & \arg\max_{\mathbf P }\mathrm{tr} (\mathbf{\Pi}_{\mathbf A(\mathbf P)} \mathbf C  \mathbf{\Pi}_{\mathbf A(\mathbf P)}) \\
\label{LS_cost}
& = & \arg\max_{\mathbf P} \mathrm{tr} (\mathbf{\Pi}_{\mathbf A(\mathbf P)} \mathbf C) 
\end{eqnarray}
where $\rm tr( \;)$ denotes the trace operator, $\mathbf C$ is the data covariance matrix $\mathbf C= \mathbf Y\mathbf Y^T$, $\mathbf I$ denotes the identity matrix.  

This is a nonlinear and nonconvex $3Q$-dimensional maximization problem. The AP algorithm \cite{AP} solves this problem by transforming it to a sequential and iterative process involving only a maximization over a single location at a time. The transformation is based on the projection-matrix decomposition formula. Let $\mathbf B$ and $\mathbf D$ be two matrices with the same number of rows, and let $\mathbf \Pi_{[\mathbf B,\mathbf D]}$ denote the projection-matrix onto the column span of the augmented matrix $[\mathbf B, \mathbf D ]$. Then:
\begin{equation}
\label{Proj_decomp}
\mathbf \Pi_{[\mathbf B,\mathbf D]}= \mathbf \Pi_{\mathbf B}+ \mathbf \Pi_{\mathbf \Pi^{\perp}_{\mathbf B}\mathbf D}, 
\end{equation}   
where $\mathbf \Pi^{\perp}_{\mathbf B}$ is the projection onto the orthogonal complement of the span of $\mathbf B$, given by $\mathbf \Pi^{\perp}_{\mathbf B} = (\mathbf I -\mathbf \Pi_{\mathbf B})$. 
 
 The AP algorithm exploits this  decomposition to transform the multidimensional maximization \eqref{LS_cost} into a sequential and iterative  process involving a maximization over only a single source at a time, with all the other sources held fixed at their pre-estimated values. More specifically, let $j+1$ denote the current iteration number, and let $q$ denote the current source to be estimated (q is sequenced from 1 to $Q$ in every iteration). The other sources are held fixed at their pre-estimated values: $\{\hat{ \mathbf p}_i^{(j+1)}\}_{i=1}^{q-1}$, which have been pre-estimated in the current iteration, and $\{\hat{ \mathbf p}_i^{(j)}\}_{i=q+1}^{Q}$, which have been pre-estimated in the previous iteration. With this notation,
 let $\mathbf A(\mathbf { \hat {P}}_{(q)}^{(j)})$ denote the $M\times (Q-1)$ matrix of the topographies corresponding to these values (note that  the $q$-th topography is excluded), given by:
\begin{equation}
\label{eq:R_i_test}
\mathbf A(\mathbf { \hat {P}}_{(q)}^{(j)})=[\mathbf l( \hat {\mathbf p}_{1}^{(j+1)}),...,\mathbf l(\hat {\mathbf p}_{q-1}^{(j+1)}),\mathbf l(\hat {\mathbf p}_{q+1}^{(j)}),...,\mathbf l(\hat {\mathbf p}_{Q}^{(j)})].
\end{equation} 
By the projection matrix decomposition \eqref{Proj_decomp}, we have:
\begin{equation}
\label{Proj_decomp_A}
\mathbf{\Pi}_{[\mathbf A(\mathbf { \hat {P}}_{(q)}^{(j)}),\mathbf l(\mathbf p_q)] }= \mathbf{\Pi}_{\mathbf A(\mathbf { \hat {P}}_{(q)}^{(j)})}+ \mathbf{\Pi}_{\mathbf {\Pi}^\perp_{\mathbf A(\mathbf { \hat {P}}_{(q)}^{(j)})}\mathbf l(\mathbf p_q)}.
\end{equation}
Substituting \eqref{Proj_decomp_A}  into \eqref{LS_cost}, and ignoring the contribution of the first term since it is not a function of $\mathbf p_q$, we get:
\begin{equation}
\label{ap_intermediate_solution}
\mathbf  {\mathbf{\hat p}}_{q}^{(j+1)} =\arg\max_{\mathbf p_q}\mathrm{tr}(  \mathbf{\Pi}_{\mathbf {\Pi}^\perp_{\mathbf A(\mathbf { \hat {P}}_{(q)}^{(j)})}\mathbf l(\mathbf p_q)}\mathbf C),
\end{equation}
or alternatively,
\begin{equation}
\label{ap_intermediate_solution}
\mathbf  {\mathbf{\hat p}}_{q}^{(j+1)} =\arg\max_{\mathbf p_q}\mathrm{tr}(  {\mathbf {\Pi}^\perp_{\mathbf Q_{(q)}^{(j)}\mathbf l(\mathbf p_q)}}\mathbf C),
\end{equation}
where
\begin{equation}
\mathbf Q_{(q)}^{(j)}= \mathbf{\Pi}_{\mathbf A(\mathbf { \hat {P}}_{(q)}^{(j)})}^{\perp}, 
\end{equation}
i.e.,  $\mathbf Q_{(q)}^{(j)}$ is a projection matrix that projects out   all but the $q$-th source at the $j$-th iteration. Using the properties of the projection and  trace operators we can rewrite \eqref{ap_intermediate_solution}  as:
\begin{equation}
\label{Iterative_solution}
\mathbf  {\mathbf{\hat p}}_{q}^{(j+1)} =\arg\max_{\mathbf p_q}\frac{\mathbf l^T(\mathbf p_q)\mathbf Q_{(q)}^{(j)}\mathbf C \mathbf Q_{(q)}^{(j)}\mathbf l(\mathbf p_q)}{{\mathbf l^T(\mathbf p_q) \mathbf Q_{(q)}^{(j)}\mathbf l(\mathbf p_q)}}.
\end{equation}
The  maximization is done by an \textit{exhaustive search} over a discrete grid representing the source space.

The initialization of the algorithm, which is critical for its good  convergence, is very straightforward. First we solve \eqref{LS_cost} for a single source, yielding
\begin{equation}
\label{eq:p_q3}
  \mathbf{\hat p}_{1}^{(0)} =\arg\max_{\mathbf p_1}\frac{\mathbf l^T(\mathbf p_1)\mathbf C \mathbf l(\mathbf p_1)}{{\mathbf l^T(\mathbf p_1) \mathbf l(\mathbf p_1)}}.
\end{equation}
Then, we  add one source at a time and solve for the $q$-th source, $q=2,...,Q$, yielding
\begin{equation}
\label{eq:p_q4}
\mathbf  {\mathbf{\hat p}}_{q}^{(0)} =\arg\max_{\mathbf p_q}\frac{\mathbf l^T(\mathbf p_q)\mathbf Q^{(0)}_{(q)}\mathbf C \mathbf Q^{(0)}_{(q)}\mathbf l(\mathbf p_q)}{{\mathbf l^T(\mathbf p_q) \mathbf Q^{(0)}_{(q)}\mathbf l(\mathbf p_q)}},
\end{equation}
where $\mathbf Q^{(0)}_{(q)}$ is the projection matrix that projects out the previously estimated $q-1$ sources: 
\begin{equation}
\mathbf Q^{(0)}_{(q)}= (\mathbf I -\mathbf{\Pi}_{\mathbf A(\mathbf { \hat {P}}^{(0)}_{(q)})}), 
\end{equation}
with $\mathbf A(\mathbf { \hat {P}}^{(0)}_{(q)})$ being the $M\times(q-1)$ matrix given by
\begin{equation}
\label{eq:R_i_test7}
\mathbf A(\mathbf { \hat {P}}_{(q)}^{(0)})=[\mathbf l( \hat {\mathbf p}_{1}^{(0)}),...,\mathbf l(\hat {\mathbf p}_{q-1}^{(0)})].
\end{equation} 
Once the initial locations of the $Q$ sources have been estimated, subsequent iterations of the algorithm, described by \eqref{Iterative_solution}, refine the estimate. The iterations continue till the localization refinement from one iteration to the next is below a pre-defined threshold. Note that the algorithm climbs the peak of \eqref{LS_cost} along lines parallel to  $\mathbf p_1,...\mathbf p_Q$, with the climb rate depending on the structure of the cost function in the proximity of the peak. Since a maximization is performed at every step, the value of the maximized function cannot decrease. As a result, the algorithm is bound to converge to a local maximum which may not necessarily be the global one. Yet, as evidenced in the simulation results,  the above  initialization procedure provides good  convergence. 

Last, it can be shown that for freely-oriented dipoles,  \eqref{Iterative_solution} and \eqref{topography_model}  lead to a generalized eigenvalue problem.

\begin{figure*}[ht!]
     \centering
     \begin{subfigure}[b]{0.3\textwidth}
         \centering
         \includegraphics[trim=0.6cm 0cm 0.0cm 0cm, clip,scale=0.43]{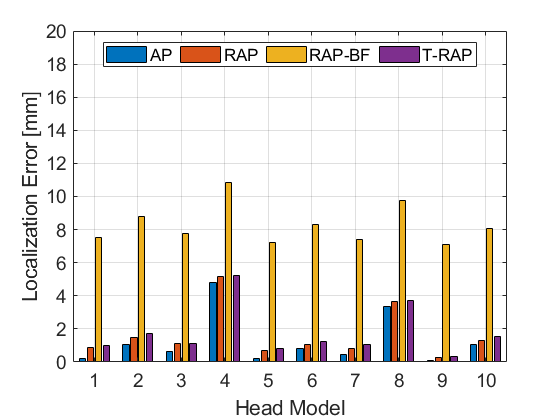}
         \caption{$\rho=0.1$}
         \label{fig:}
     \end{subfigure}
     \hfill
     \begin{subfigure}[b]{0.3\textwidth}
         \centering
         \includegraphics[trim=1.0cm 0cm 1.25cm 0cm, clip,scale=0.43]{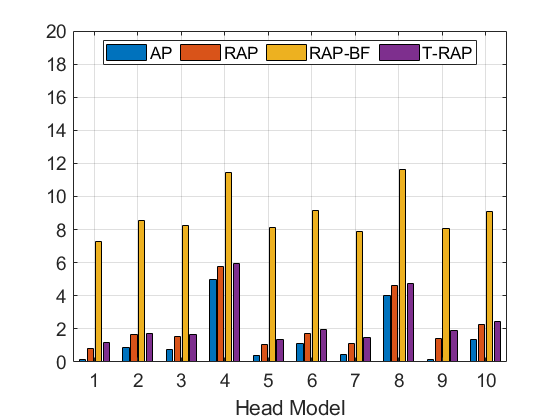}
         \caption{$\rho=0.5$}
         \label{fig:}
     \end{subfigure}
     \hfill
     \begin{subfigure}[b]{0.3\textwidth}
         \centering
         \includegraphics[trim=1.25cm 0cm 1.2cm 0cm, clip,scale=0.43]{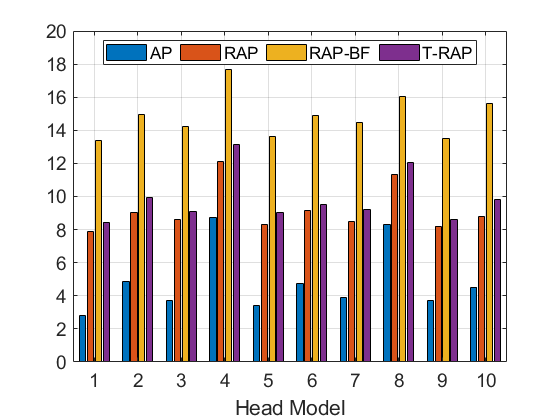}
         \caption{$\rho=0.9$}
         \label{fig:}
     \end{subfigure}
        \caption{Localization error  with varying inter-source correlation ($\rho$) in the presence of head model registration errors:  (1) 1 mm x-axis translation posterior.  (2) 2 mm x-axis translation posterior. (3) $1 ^{\circ}$ x-axis rotation right tilt. (4) $2 ^{\circ}$ x-axis rotation right tilt. (5) 1 mm z-axis translation upward. (6) 2 mm z-axis translation upward. (7) $1 ^{\circ}$ y-axis rotation upward. (8) $2 ^{\circ}$ y-axis rotation upward. (9) 1 mm y-axis translation right.  (10) 2 mm y-axis translation right.}
        \label{fig:three graphs}
\end{figure*}
\vskip -2.5pt
\section{Performance evaluation}
\vskip -7pt
\subsection{Simulated Data}
\vskip -2.5pt
We assessed the performance of the AP localization method against the RAP-beamformer (RAP-BF) \cite{ilmoniemi2019brain}, RAP-MUSIC (RAP) \cite{Mosher1999} and Truncated RAP-MUSIC (T-RAP) \cite{makela_truncated_2018}. Performance evaluation relied on Monte-Carlo simulations (1000 random samples per condition) of pairs of sources with different levels of inter-source correlation. The geometry of the sensor array was based on the 306-channel Megin Triux MEG device. The geometry of the MEG source space was modeled with the cortical manifold extracted from an adult human subject's MR data using Freesurfer \cite{fischl_2004}. Sources were restricted to 15,002 grid points over the cortex. The lead field matrix was estimated using BrainStorm \cite{Brainstorm2011} based on the overlapping spheres head model \cite{Huang1999}. Gaussian white noise was generated and added to the MEG sensors to model instrumentation noise at 0 dB SNR level. Source time courses were modeled with 50 time points sampled as mixtures of sinusoidal signals with frequencies randomly chosen between 10 Hz to 30 Hz, following the protocol of \cite{makela_truncated_2018}.

The AP method had the lowest localization error across all inter-source correlation levels and forward model errors (Figure \ref{fig:three graphs}). Specifically, the average localization error was 10 mm for AP, 13.77 mm for RAP, 17.7 mm for RAP-BF, and 13.76 mm for T-RAP. In all methods, localization error was highest for $\rho=0.9$ inter-source correlation.
\subsection{Multimodal Sensory Human Data}
\vskip -2.5pt
\begin{figure*}[ht!]
    \centering
    \includegraphics[scale=0.37]{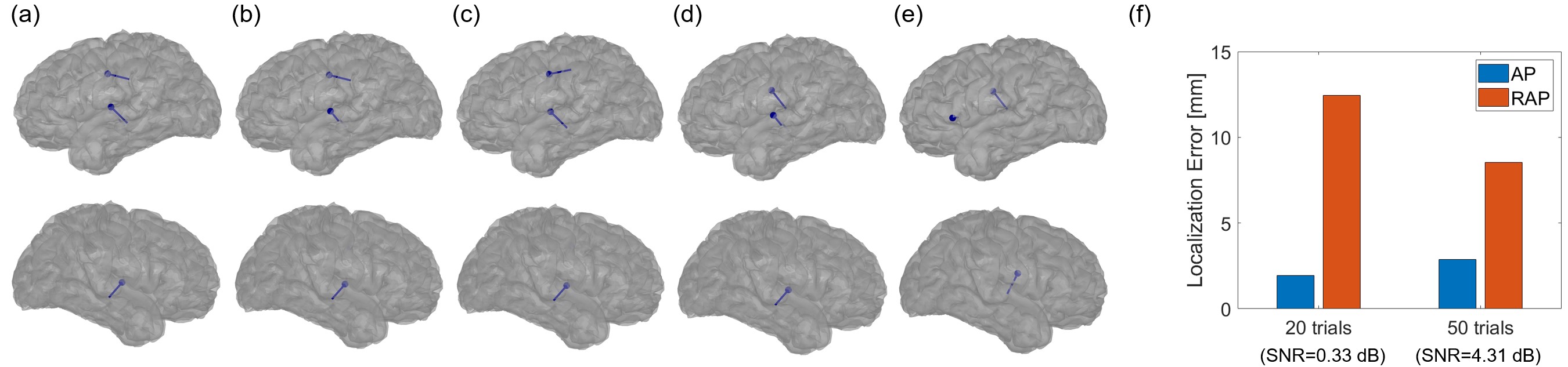}
    \caption{Localization performance for multisensory MEG human data. (a) Ground truth dipoles from a human experiment presenting somatosensory and auditory stimuli. Somatosensory and auditory evoked responses were then combined with different numbers of trials to assess localization performance in multisensory data: (b,c) AP method with 50 and 20 trials, respectively; (d,e) RAP-MUSIC with 50 and 20 trials, respectively. Results are shown in left and right lateral views of semi-transparent cortex. (f) Localization error averaged across the three dipoles.}
    \label{fig:realhumandata}
\end{figure*}
We collected MEG data from a human participant in a multimodal sensory experiment that presented somatosensory stimuli with electrical stimulation to the right median nerve, and auditory stimuli with tubal-insert earphones to both ears. Fifty trials were recorded for each stimulus type with an interstimulus interval 500ms for the somatosensory stimuli and 1200ms for the auditory stimuli. The data were collected with a MEGIN Triux MEG device. Sources were restricted to 15,002 grid points over the subject-specific cortex extracted with Freesurfer \cite{fischl_2004}. The lead field matrix was estimated using the overlapping spheres head model \cite{Huang1999} of the BrainStorm software \cite{Brainstorm2011}. Raw data was pre-processed with the Maxfilter software (MEGIN, Helsinki) with default parameters to reduce noise with spatiotemporal filters \cite{Taulu_2006}, and whitened using a regularized noise covariance matrix.

To establish ground truth, we fit single dipoles separately to the somatosensory and auditory data (all tested localization methods are equivalent for the single dipole case). The dipoles were localized to well-known areas in the left primary somatosensory cortex and bilateral primary auditory cortex, respectively. In the auditory case, we fit data separately for the  left and right sensors to estimate a dipole in the left and right primary auditory cortex, respectively (Figure \ref{fig:realhumandata}a).

The somatosensory and auditory data were then combined, by adding the corresponding sensor measurements at 100 ms and 40 ms respectively (the time of peak response), to yield multimodal data. Data were combined with 50 trials (SNR = 4.31 dB) or 20 trials (SNR = 0.33 dB) to assess the performance of the AP and RAP-MUSIC methods at different SNR levels. 

The AP method localized the three dipoles in close agreement with the ground truth (Fig. \ref{fig:realhumandata}b,c,f). In contrast, the RAP-MUSIC method yielded approximate results and was highly variable with the number of trials (Fig. \ref{fig:realhumandata}d,e,f).

\section{Conclusions} 
We have presented a new sequential and iterative solution to the localization of MEG/EEG signals based on minimization of the LS criterion by the AP algorithm. Our simulation and real data results demonstrated that the AP algorithm performs robustly  for multiple sources with wide range of inter-source correlation values. Taken together, our work demonstrated the high performance of the AP algorithm in localizing MEG sources. It also revealed the importance of iterating through all sources multiple times until convergence instead of recursively scanning for sources and terminating the scan after the last source is found, which is the approach typically adopted by existing scanning methods. The results show the clear superiority of the AP solution in the presence of forward model errors resulting from head rotations and translations.

The analysis tools used in the current study are
available at  \url{https://alternatingprojection.github.io/}.

\section{Compliance with Ethical Standards}
\vskip -5pt
This study was performed in line with the principles of the Declaration of Helsinki. The human participant gave a written informed consent, and the study was approved by the IRB of the Massachusetts Institute of Technology.
\vspace{-7pt}

\section{Acknowledgments}
\vskip -3pt
This work was supported by the United States-Israel Binational Science Foundation grant 2020805 to A.A. and NIH grant 1R01EY033638-01 to D.P.. The authors have no relevant personal financial or non-financial interests to disclose.

\bibliographystyle{IEEEbib}
\vspace{-3pt}
\bibliography{references}

\begin{thebibliography}{10}

\bibitem{ilmoniemi2019brain}
R.J. Ilmoniemi and J.~Sarvas,
\newblock {\em Brain Signals: Physics and Mathematics of MEG and EEG},
\newblock MIT Press, 2019.

\bibitem{mosher_multiple_1992}
J.C. Mosher, P.S. Lewis, and R.M. Leahy,
\newblock ``Multiple dipole modeling and localization from spatio-temporal
  {MEG} data,''
\newblock {\em IEEE Transactions on Biomedical Engineering}, vol. 39, no. 6,
  pp. 541--557, June 1992.

\bibitem{hamalainen_magnetoencephalographytheory_1993}
Matti Hämäläinen, Riitta Hari, Risto~J. Ilmoniemi, Jukka Knuutila, and
  Olli~V. Lounasmaa,
\newblock ``Magnetoencephalography—theory, instrumentation, and applications
  to noninvasive studies of the working human brain,''
\newblock {\em Reviews of Modern Physics}, vol. 65, no. 2, pp. 413--497, Apr.
  1993.

\bibitem{uutela_global_1998}
K.~Uutela, M.~Hamalainen, and R.~Salmelin,
\newblock ``Global optimization in the localization of neuromagnetic sources,''
\newblock {\em IEEE Transactions on Biomedical Engineering}, vol. 45, no. 6,
  pp. 716--723, June 1998.

\bibitem{jiang_comparative_2003}
Tianzi Jiang, An~Luo, Xiaodong Li, and F.~Kruggel,
\newblock ``A {Comparative} {Study} {Of} {Global} {Optimization} {Approaches}
  {To} {Meg} {Source} {Localization},''
\newblock {\em International Journal of Computer Mathematics}, vol. 80, no. 3,
  pp. 305--324, Mar. 2003.

\bibitem{darvas_mapping_2004}
F.~Darvas, D.~Pantazis, E.~Kucukaltun-Yildirim, and R.M. Leahy,
\newblock ``Mapping human brain function with {MEG} and {EEG}: methods and
  validation,''
\newblock {\em NeuroImage}, vol. 23, pp. S289--S299, Jan. 2004.

\bibitem{LCMV}
{B. D. van Veen, W. van Drongelen, M. Yuchtman, and A. Suzuki},
\newblock ``Localization of brain electrical activity via linearly constrained
  minimum variance spatial filtering,''
\newblock {\em IEEE Trans. Biomed. Eng.}, vol. 44, no. 9, pp. 867--880, Sep
  1997.

\bibitem{VerbaRobinson}
J.~Vrba and S.~E. Robinson,
\newblock ``Signal processing in magnetoencephalography,''
\newblock {\em Methods}, vol. 25, pp. 249--271, 2001.

\bibitem{hui_identifying_2010}
Hua~Brian Hui, Dimitrios Pantazis, Steven~L. Bressler, and Richard~M. Leahy,
\newblock ``Identifying true cortical interactions in {MEG} using the nulling
  beamformer,''
\newblock {\em NeuroImage}, vol. 49, no. 4, pp. 3161--3174, Feb. 2010.

\bibitem{CorrBF}
{M. J. {Brookes}, C. M. {Stevenson}, G. R. {Barnes}, A. {Hillebrand}, M. I.
  {Simpson}, S. T. {Francis}, and P. G. {Morris}},
\newblock ``Beamformer reconstruction of correlated sources using a modified
  source,''
\newblock {\em NeuroImage}, vol. 34, no. 4, pp. 1454--1465, 2007.

\bibitem{DCBF}
{M. Diwakar and M.-X Huang and R. Srinivasan and D. L. Harrington and A. Robb
  and A. Angeles and L. Muzzatti and R. Pakdaman and T. Song, R. J. Theilmann
  and R. R. Lee },
\newblock ``Dual-core beamformer for obtaining highly correlated neuronal
  networks in meg,''
\newblock {\em NeuroImage}, vol. 54, no. 1, pp. 253--263, 2011.

\bibitem{EDCBF}
M.~Diwakar, O.~Tal, T.~Liu, D.~Harringtona, R.~Srinivasan, L.~Muzzatti,
  T.~Song, R.~Theilmann, R.~Lee, and M.-X. Huang,
\newblock ``Accurate reconstruction of temporal correlation for neuronal
  sources using the enhanced dual-core meg beamformer,''
\newblock {\em NeuroImage}, vol. 56, pp. 1918--1928, 2011.

\bibitem{Mosher1999}
J.~C. {Mosher} and R.~M. {Leahy},
\newblock ``Source localization using recursively applied and projected {(RAP)
  MUSIC},''
\newblock {\em IEEE Transactions on Signal Processing}, vol. 47, no. 2, pp.
  332--340, Feb 1999.

\bibitem{makela_truncated_2018}
Niko Mäkelä, Matti Stenroos, Jukka Sarvas, and Risto~J. Ilmoniemi,
\newblock ``Truncated {RAP}-{MUSIC} ({TRAP}-{MUSIC}) for {MEG} and {EEG} source
  localization,''
\newblock {\em NeuroImage}, vol. 167, pp. 73--83, Feb. 2018.

\bibitem{SPMBrainMap}
S.~{Baillet}, J.~C. {Mosher}, and R.~M. {Leahy},
\newblock ``Electromagnetic brain mapping,''
\newblock {\em IEEE Signal Processing Magazine}, vol. 18, no. 6, pp. 14--30,
  Nov 2001.

\bibitem{AP}
I.~{Ziskind} and M.~{Wax},
\newblock ``Maximum likelihood localization of multiple sources by alternating
  projection,''
\newblock {\em IEEE Transactions on Acoustics, Speech, and Signal Processing},
  vol. 36, no. 10, pp. 1553--1560, Oct 1988.

\bibitem{fischl_2004}
Bruce Fischl, David~H. Salat, André~J.W. van~der Kouwe, Nikos Makris, Florent
  Ségonne, Brian~T. Quinn, and Anders~M. Dale,
\newblock ``Sequence-independent segmentation of magnetic resonance images,''
\newblock {\em NeuroImage}, vol. 23, pp. S69--S84, Jan. 2004.

\bibitem{Brainstorm2011}
François Tadel, Sylvain Baillet, John~C. Mosher, Dimitrios Pantazis, and
  Richard~M. Leahy,
\newblock ``Brainstorm: a user-friendly application for {MEG/EEG} analysis,''
\newblock {\em Computational intelligence and neuroscience}, 2011.

\bibitem{Huang1999}
M.~X. Huang, J.~C. Mosher, and R.~M. Leahy,
\newblock ``A sensor-weighted overlapping-sphere head model and exhaustive head
  model comparison for {MEG},''
\newblock {\em Physics in medicine and biology}, vol. 44, pp. 423–440, 1999.

\bibitem{Taulu_2006}
S~Taulu and J~Simola,
\newblock ``Spatiotemporal signal space separation method for rejecting nearby
  interference in {MEG} measurements,''
\newblock {\em Physics in Medicine and Biology}, vol. 51, no. 7, pp.
  1759--1768, mar 2006.

\end{thebibliography}
\vskip -3pt

\end{document}